\documentclass[runningheads]{svmult}

\usepackage{makeidx}   
\usepackage{graphicx}  
\usepackage{subeqnar}  
\usepackage{multicol}  
\usepackage{physprbb}  
\makeindex             

\begin{document}
\title*{Nature, Nurture or Not Sure?\protect\newline 
A Debate About SGRs and AXPs}
\toctitle{Nature, Nurture or Not Sure?\protect\newline 
A Debate About SGRs and AXPs}
%
%
\titlerunning{Nature, Nurture or Not Sure?}
%
\author{R.C. Duncan}
\authorrunning{R.C. Duncan}
%
%
\institute{Dept.~of Astronomy, University of Texas, Austin, Texas 78712 USA}

\maketitle              

\begin{abstract}
Marsden, Lingenfelter, Rothschild \& Higdon have given 
arguments against the magnetar model for Soft Gamma Repeaters (SGRs)
and Anomalous X-ray Pulsars (AXPs), as forcefully advocated 
by R. Rothschild at this meeting. We critique these arguments, showing: (1) 
The claim that SGRs and AXPs are born in unusually dense regions of the
ISM is not supported in any compelling way by observations. (2) Even if 
this claim were true, it would not argue against the magnetar model. 
Moreover, all observations can be accounted for if magnetars 
have shorter observable lifetimes than do radiopulsars, 
in agreement with theoretical expectations, but no systematically
different ambient ISM densities. 
(3) The suggestion that accretion  
onto the neutron star is directly influenced by the ISM in a way that
explains the difference between SGRs/AXPs and radiopulsars, is not 
possible.  The mass inflow rate during later stages of supernova
remnant expansion when accretion can be 
influenced by backpressure from the ISM is many orders of magnitude 
too small.   (4)  Accretion-based models are unable to account 
for the hyper-Eddington bursts and flares which are the defining characteristic
of SGRs.  (5) Accretion disk models also predict optical and IR emission 
with higher luminosities than are observed. 
\end{abstract}

\section{Introduction}

Marsden {\it et al.}~present observational evidence, 
culled from the literature, to show that SGRs and AXPs are associated with 
supernova
remnants (SNRs) expanding into unusually dense parts of the interstellar
medium (ISM).  They claim that this disproves the magnetar
hypothesis \cite{m1}:  
``$\ldots$the environments surrounding SGRs and AXPs are significantly
different than `normal' neutron stars in a way which is \emph{inconsistent}
with the hypothesis that the properties of these sources derive solely
from an  innate characteristic such as a superstrong magnetic field."

Marsden {\it et al.}~go on to claim that everything about SGRs and
AXPs, including their bursting behavior, can be explained if these objects 
are $B \sim 10^{12}$ G neutron stars accreting from SNRs in
high-density regions of  the ISM.  

Three different version of the Marsden {\it et al.}~paper were widely 
distributed and posted on the astro-ph archives \cite{m1,m2,m3}. \ We will 
discuss arguments from all three versions, since R. Rothschild gave 
arguments from all three versions at this meeting.
(See Rothschild's contribution to this volume).  

In \S 2 of this paper, we show that most of the associations of 
SGRs and AXPs with SNRs claimed 
by Marsden {\it et al.}~are subject to reasonable doubt.  Moreover, 
the SNR remnant ages have very large uncertainties, spanning more than an 
order of magnitude, if one assumes no {\it a priori} knowledge of the 
ISM ambient density.  This means that the value of the ambient ISM density 
is not strongly constrained by SNR measurements, even if all 
the proposed SNR associations are assumed to be valid. 
We conclude that there is no persuasive evidence for the hypothesis 
that SGRs and AXPs are born in regions of the ISM with unusually 
high density.

  In \S 3 we show that, even if it \emph{was} proven that SGRs and AXPs 
are born in denser ISM environments than radiopulsars, this would not 
necessarily argue against 
the magnetar model.  We also offer an alternative interpretation of the data. 
Although the data set of Marsden {\it et al.}~does not unambiguously constrain 
ambient ISM densities, it does give some evidence  
that observed SGRs/AXPs are associated with more compact (smaller-radius) 
SNRs than are radiopulsars on average. This can be naturally 
understood if SGRs/AXPs have comparatively short observable lifetimes, 
as expected in the magnetar model.  

 In \S 4 we turn to a scenario discussed by R. Rothschild
at this meeting, namely that accretion
onto a neutron star is directly influenced by a dense ambient ISM in a way 
that can explain all differences between SGR/AXPs and radiopulsars.
The rate of accretion onto a high-velocity neutron star 
from  a SNR in the late stages when it experiences significant backpressure 
from the ISM is quantified.   The resulting values of $\dot{M}$ are 
too small by many orders of
magnitude to drive the  stellar spindown or the X-ray emissions of SGRs
and AXPs.  Thus the answer to  ``Nature or Nurture?" question is 
clearly ``Nature." 

This alone does not necessarily mean that the stars are magnetars.
Alternatives to the magnetar model which posit accretion from  a
``fallback disk" which is formed at very early times within the SNR from 
co-moving ejecta---when the immediate environment of the neutron star
is  not directly influenced by the surrounding ISM (again, not
``Nurture")---are  also possible.  This scenario was proposed by Chatterjee,
Herquist and Narayan and  by other authors \cite{chatter}.  

In the concluding section (\S 5) we briefly summarize some general 
arguments against accretion-powered models of SGRs an AXPs. This evidence 
seems to exclude the theoretically reasonable ``fallback disk" models as well 
as the more dubious ``pushback disk" models suggested by Marsden {\it et al.} 

\section{Observational Evidence for a Dense ISM around SGRs/AXPs?}

The method of Marsden {\it et al.}~is as follows.  They begin by 
adopting (in most cases, dubiously) a particular SNR association for
each SGR and AXP.  Given the SNR's measured angular size, $\theta_{snr}$,
published estimates of the distance $D$ to each object imply a SNR 
radius $R= \theta_{snr} D$. \ If the age $t$ of the SNR is 
known, then the density into which the supernova expands can be
estimated via the Sedov solution, which in c.g.s. units is 
$\rho = 1.17 \,E \,  R^{-5} \, t^2 $. 
Figure 3 in ref.~\cite{m3} shows SGR and AXP points plotted on 
the $R - t$ plane, along with contours of constant
ambient density, defined according to the Sedov formula where one
assumes $E = 10^{51}$ erg. 
A similar process is done for radiopulsars (Fig.~4),  
leading to the claim that the two sets of data are significantly
discrepant, with radiopulsars in lower-density regions of the plane.   

This is dubious.  First of all, the $R^{-5}$ dependence of $\rho$ implies 
a $D^{-5}$ dependence on the inferred density.  If $D$ goes up by 2, $\rho$ 
goes down by 30.  Many or most of the $D$ values are uncertain 
by factors of 2.   This is true even when the measurement uncertainty is 
as small as $10\%$, as quoted by Corbel {\it et al.}~in their elaborate 
arguments for two SGR distances \cite{corbel1,corbel2}, because systematic 
uncertainties in data interpretation can exceed measurement
uncertainties. It is notoriously difficult to unambiguously determine 
the distances of molecular clouds along lines of sight extending 
across the Galactic disk.

Furthermore, the ages $t$ are very poorly constrained. 
The usual way to determine the age of a SNR is to assume something about
the ambient density, and use the Sedov solution. Here this cannot
be done since the density is the quantity that we ultimately want to determine.
For most objects, Marsden {\it et al.}~adopt a lower limit on the age 
given by min$\{t_{fe},t_v\}$ , where
$t_{fe} =  D_{min}  \theta_{snr} / v_{ej}$,  and
$t_v = D_{min} \theta_* / v_{max}$. In these equations, 
$v_{ej} = 10,000$ km s$^{-1}$ is the assumed ejecta velocity, 
$\theta_*$ is the angular displacement of the star
from the SNR center, and $v_{max} \equiv 2000$ km s$^{-1}$.
But in every case $t_v$ is greater than $t_{fe}$, so $t_v$ is effectively the
lower limit. This is the time that a star moving at a transverse velocity
$v_{max}$ would move from the center of the SNR to its present location,
at a distance $D_{min}$, the minimum plausible value of $D$.
But the center of the SNR can only be roughly fixed since the SNRs 
generally show deviations from sphericity and inhomogeneities.  Moreover,
the choice of 2000 km s$^{-1}$ is arbitrary;  larger velocities cannot
be excluded.  The logic of using an X-ray star's assumed velocity 
to infer a lower bound to the density of the environment seems convoluted. 
The estimate $t_{fe}$, never applied, would have the error bars extend much
further to the right, to the \emph{Free Expansion} line in Fig.~3 at
$D = D_{min}$. This is a believable lower bound;
it shows that the possible ages span orders of magnitude.
The upper limits of 3 kyr for all but three points are also essentially
arbitrary:  ``we choose 30 kyr, which is the maximum estimated age [value
of $(P/ 2 \dot{P})$ ] for the
SNR/radiopulsar associations in Table 3."   
Why is it a reasonable to use the maximum age of 
a completely distinct set of objects?

  In the three cases where detailed modeling of the SNR has been done, 
the error box is chosen to span the range of published values from models.  
But this
underestimates the uncertainty also: each estimate really has an 
uncertainty itself, based on the data and methods used.  This might not 
quoted, since it is hard to estimate; but there is no reason to think
that it is smaller than the range of quoted ages, at least not without 
studying the issue.  

    By plotting each point at the arithmatic (not logarithmic) 
center of an error range that spans 
more than an order of magnitude (see Fig.~3  of ref.~\cite{m3}), one fosters 
the impression that the density lies at the high end of the (claimed) 
permitted range.  The true uncertainties are certainly not Gaussian 
or even log-Gaussian. Thus the data should really be plotted as error 
boxes, each extending horizontally across most of the plot. This would 
convey the fact that we have almost no information about the ages of these 
objects that is independent of assumptions about ambient density. 
The possible values of ambient density inferred from the SNRs also 
span orders of magnitude.

  Note that for radio pulsars (Fig.~4), Marsden {\it et al.}~say:  
``The MDR timing ages
of radio pulsars, unlike those for SGRs and AXPs, are thought to be good
measures of their true ages."   However, Gaensler and Frail, 
ref.~\cite{gf2000}, have shown that the true age of PSR B1757-24 exceeds 
39,000 yrs, and has most probable value of 170,000 yrs, which is more 
than 10 times its MDR age, $t_{mdr} \equiv P/2 \dot{P}$.   This suggests 
that the MDR ages used in Fig.~4 are lower 
bounds, and the inferred densities are thus lower bounds on the true 
densities. 

Finally, many of the adopted SNR associations are themselves dubious.  
Among AXPs, only 1E 1841-045, 1E 2259+586, and J1845-0258 have 
well-established SNR associations, and it is not certain that the third
object is an AXP since its $\dot{P}$ is not yet measured. 
For the SGRs, all SNR associations have been questioned in the
literature (e.g., ref.~\cite{gaensler}). 
The claimed association of SGR 1801-23 is especially dubious, since 
its error box is a 3.5 degree long annulus intersecting several candidate
SNRs \cite{cline}.  Most SGRs are located near the edge of the 
putative associated SNR, requiring very
high-velocity neutron stars $V_{sgr}> 1000$ km s$^{-1}$.   
Note that the radio emission around SGR 1806-20 may not be SNR at 
all, but a synchrotron nebula fed by a LBV star \cite{hurley}.

\section{Comparisons with the magnetar model}

Even if SGRs and AXPs \emph{did} tend to come from denser regions of the ISM 
than do radiopulsars, this would not argue against the magnetar
hypothesis.

Magnetars are thought to form during stellar core collapse events when the  
proto-neutron star is rotating rapidly enough during its convective
phase, just after collapse, to support an $\alpha$--$\Omega$ 
dynamo \cite{dt92,dt96}. 
The progenitor objects may be massive stars which, for one reason or 
another, retain substantial angular momentum and do not 
transport too much angular momentum outward from the core material 
before collapse.
Complex issues of star formation and rotational
evolution (including possible binary interactions) may be involved 
in producing this subset of objects.
Although these complexities are not fully understood,  
it is plausible that the magnetar-producing stars tend to
supernova  in regions of the ISM with somewhat different properties
than radiopulsar-producing stars do. 
For example, if magnetars tended to 
result from more massive progenetor stars than did radiopulsars, 
they would be found preferentially among the first 
generation of supernovae in star-forming regions, before a hot 
superbubble forms.  This kind of effect could explain a 
correlation with ambient density as claimed by Marsden {\it et al.} 

 We favor a different interpretation of the data.
 In the absence of independent information about the ages of
the SGRs/AXPs, what Marsden {\it et al.}~really have studied is simply 
\emph{the distribution of SNR radii $R$,} as given by 
the projection of points in Fig.~3 of ref.~\cite{m3} on the vertical 
($R$) axis.  A comparison of this distribution with the analogous 
distribution for Marsden {\it et al.}'s sample of radiopulsars 
with $t_{mdr} < 30,000$ yr.  (Fig.~4; projected on vertical axis)
suggests that \emph{SGRs/AXPs tend to be associated with smaller-$R$ SNR 
than do young radiopulsars.} (Of course, this conclusion is only
correct if we accept the dubious SNR--star associations.)

This could be explained by various selection effects which operate
on both sets of objects. In particular, it is likely that 
\emph{radiopulsars are observable for significantly longer periods of time 
than are SGR/AXPs.} This fact alone could account for the trend 
in SNR sizes.
 
In the magnetar model, all observable magnetically-powered emissions are
likely to shut off when the liquid interior of the star cools 
sufficiently to halt ambipolar diffusion. 
Thompson and I showed that a magnetar's core temperature declines 
only weakly with stellar age, $T \sim t^{-1/7}$, at early times when 
ambipolar diffusive heating is balanced by neutrino cooling 
(eq.~[36] in ref.~\cite{td96}). This weak power-law decline turns down 
sharply when neutrino cooling becomes dominated by rapid photon cooling
from the star's surface, essentially causing interior 
magnetic dissipative heating and core magnetic evolution to cease. 
We originally  estimated that this occurs at an age 
$t\sim 10^6$ yrs (eq.~[75] in ref.~\cite{td96}). 
But if a magnetar has a light-element envelope, then surface photon cooling
is greatly accelerated (ref.~\cite{hh98}), and it dominates the thermal 
history sooner, freezing out ambipolar diffusion.  A magnetar could then 
become X-ray dark and burst--quiet as early as $\sim 10^4$ or $10^5$ years
(C. Thompson, private communication) with a strong magnetic field 
$B_{core}> 10^{14}$ G
frozen in its core thereafter, decaying only on the enormously long time-scales
of ohmic diffusion.  This ``early death" could account for the narrow 
observed period range of X-ray bright SGRs and AXPs, due to observational
selection effects in both steady X-ray emissions and bursts.  No true 
decay of $B_{dipole}$ must be invoked (cf.~ref.~\cite{colpi00}). 

In the case of radiopulsars, the lifetime to spin down past 
the radio death line undoubtedly exceeds the time for
SNR fading and dispersal, by several orders of magnitude.  
If some magnetars become undetectable 
before their SNRs do, then radiopulsars would tend to have 
older and larger SNRs than (observed) magnetars, even if the ambient 
density and SNR evolutionary tracks were identical for the two classes 
of neutron stars. 

We conclude this section with a note about implications of this 
picture. Because magnetar activity is ephemeral,
slowly-rotating {\it dead magnetars} with large magnetic dipole fields
should be common in the Galaxy.  Indeed there could be $> 10^7$ of them if
a substantial fraction of all neutron stars are magnetars, as suggested
by studies of all neutron stars known to be associated with young
SNRs \cite{kaspi}.  Detecting dead magnetars is a challenging problem 
for astronomers.

\section{Propeller-driven spindown in a dense ISM?}

Marsden {\it et al.}~propose an alternative theory: that interaction of  a
$10^{12}$ G neutron star with its SNR in a relatively dense ISM can
explain  everything about SGRs and AXPs.  

Accretion by a compact star moving through a gaseous medium is a
classical problem in astrophysics, first quantified by Bondi, 
Hoyle and Lyttleton.  Here we only consider accretion that occurs after the
ISM can influence the immediate environment of the star; i.e., accretion 
occurring after the reverse shock reaches the high-velocity neutron
star at $t >10^2$ years.

Quantitative estimates of this were not given in the first version 
of the Marsden {\it et al.}~paper \cite{m1}. The second version \cite{m2} did 
include some 
estimates. Careful scrutiny of ref.~\cite{m2} reveals that, even assuming 
extremely unlikely and optimistic values of all parameters, accretion 
influenced by the ISM fails by many orders of magnitude to produce 
enough inflowing material to either spin down the star in $\sim 10^4$ years or 
to power the observed X-ray luminosities of SGRs and AXPs.  The third version 
of the paper \cite{m3} omits the analysis again.

 It is not surprising that such ``pushback accretion" does not work.  
A supernova is a powerful explosion and even ``dense" parts of ISM 
(ambient density assumed to be $\sim 10$--$10^2$ gm cm$^{-3}$; 
ref.~\cite{m3}) are 
superb vacua by laboratory standards. Thus the pushback
phenomena is not very strong.
Here we use some rough approximations to quantify this.  More reliable 
analysis would require the use of detailed SNR simulations. 

Since the ejecta expands nearly homologously with uniform density and nearly
constant surface velocity during the initial free-expansion 
phase (e.g., ref.~\cite{longair}), the
ejecta is almost comoving with the star at early times after the star
moves away from the center of the SNR.  This is the epoch when  a
fallback (not pushback) disk might form, see e.g., 
ref.~\cite{chatter,chevalier}.

During this period the surrounding interstellar medium (ISM) has 
essentially no influence on the immediate environment and accretion
rate  of the neutron star.  The time scale for the neutron star to 
intercept the reverse shock and first experience possible 
influence  of the surrounding ISM, is $> 10^2$ years.
By this time, the density of the ejecta declines to roughly
$n(t) \simeq 10 M_\odot /[ (4\pi /3) (V_F \, t)^3 ]$ or
 \hbox{$n(t)\simeq 10^2 \, t_2^{-3} \, V_4^{-3}$ cm$^{-3}$,} if we
optimistically scale to $t = 10^2 \, t_2$ yrs.
The density of the shell material is in the range between 4 times
this number (on the shell side of the reverse shock) and 4 times the
ISM density (on the shell side of the outer shock).  This factor 4 is
of course the strong-shock enhancement factor for $\Gamma =5/3$.

  The accretion rate is
$\dot{M} = (2GM)^2 \rho/V^3$ where  $V = [(kT/m) + V_*^2]^{1/2}$,
for a neutron star traveling through gas of temperature $T$ with   
a stellar velocity relative to the gas  $V_*$  
(e.g., ref.~\cite{shapteuk})
 For  $M = 1.4 \, M_\odot$,  $V_* = 10^3 \, V_3$ km s$^{-1}$  and cold gas with 
density  $\rho = m n$, where  $n =  10^2 \, n_2$ cm$^{-3}$, one finds
\hbox{$\dot{M} = 7 \times 10^7 \, n_2 \, V_3^{-3}$ gm s$^{-1}$.} It is 
less for hot gas.  In $10^4 \, t_4$ years, as the star crosses the SNR
shell, this would accumulate only $10^{-14} \, n_2 \, t_4 \, V_3^{-3} M_\odot$,
even if the density declined no further and the gas were not shock-heated. 

The $\dot{M}$ needed to drive propeller spindown with 
$( \dot{P} / P ) = 1 \times 10^{-11}$, as found for SGRs \cite{kouv}, 
is $\dot{M} = 4 \times 10^{16}  \, B_{12}^{-8/3}  \, 
(\dot{P}/[P \times 10^{-11}])^{7/3}$
gm s$^{-1}$ (e.g., ref.~\cite{shapteuk}). 
Note that this is somewhat greater (by $\sim 10^2$) than the 
$\dot{M}$ needed to power X-ray emissions with luminosities 
$L_x \sim 10^{35}$ erg s$^{-1}$ as observed, assuming accretion 
down to the surface.

Comparing the above two values of $\dot{M}$, it is evident that \emph{accretion
influenced by the ISM fails by a factor of $\sim 10^9$ for these scalings.} \ 
To get the needed $\dot{M}$, one must have $V_* < 1.2 \, n_2^{1/3} \, 
B_{12}^{8/9}$ km s$^{-1}$.
Thus one needs for the gas in the SNR shell to be nearly co-moving 
\emph{and} it also must be very cold, with $T < 180$ K.  Neither
condition is satisfied, since the shell material is significantly
decelerated as soon as it crosses the reverse shock, slowing to
$[(2/5) R/t]$ in the Sedov phase,  and it is also shock-heated to
temperatures $\sim 10^6$ K or more,  as evinced by the X-rays it emits.   
Once the
star passes outside the  shell, the ISM may be cool, but it is far
from co-moving.  A shortfall in $\dot{M}$ by a very large factor
is unavoidable.

The presence of some dust in the cooling SNR, and nonspherical gas flows,
do not save the scenario.
Insofar  as dust grains accrete like a zero-pressure, collisionless
gas, they will have much smaller $\dot{M}$, by a  
factor $(R_{NS} V^2/2GM)$, where $R_{NS}$ is the neutron star's radius
(e.g., \S 14.2 in ref.~\cite{shapteuk}).  More realistically,
dust will be swept along with the gas and have little effect on $\dot{M}$. 
As for non-spherical flows, the vast expansion factors in the SNR damp
out any velocity variations  except for Rayleigh-Taylor instabilities
near the reverse shock, but these  transient motions contain less than
$1\%$ of the SNR kinetic energy and occur  in a narrow mass shell. 
These effects cannot make up  the enormous $\dot{M}$ 
discrepancies quoted above.

\section{Other problems with accretion-powered models}

  It is not known how accretion-powered models 
can account for the hyper-Eddington bursts and flares which are the 
hallmark of SGRs.  The hard-gamma initial spikes of giant flares with 
peak luminosities $\sim 10^7 \, L_{\rm Edd}$, lasting several tenths of 
a second, are especially difficult to account for in accretion-powered 
models.  However, the observed properties of SGR outbursts are
consistent  with magnetically-driven instabilities on magnetars  
\cite{td95,feroci,gogus,td01}.

 Furthermore, if SGRs and AXPs have accretion disks then they should 
emit optical and infrared radiation, due to the reprocessing of X-rays and 
viscous dissipation within the disk itself. 
The luminosity of this emission is not sensitive to the disk composition, 
as long as the disk is optically
thick. Observations of optical and IR emission from several SGRs and AXPs 
(e.g.~ref.~\cite{hulleman1,hulleman2}) find luminosities far below the 
predictions of disk models (e.g.~ref.~\cite{perna}).

\end{document}